\documentclass{optica-article}

\journal{opticajournal} 

\articletype{Research Article}

\usepackage{listings}
\usepackage{xcolor}
\usepackage{siunitx}
\usepackage{lineno}
\newcommand{\diff}{\mathop{}\!\mathrm{d}}  
\usepackage{bm}

\begin{document}
\lstdefinestyle{MATLABStyle}{
    language=Matlab,
    basicstyle=\ttfamily\footnotesize,
    keywordstyle=\color{blue},
    commentstyle=\color{green},
    stringstyle=\color{red},
    numbers=left,
    numberstyle=\tiny\color{gray},
    stepnumber=1,
    numbersep=10pt,
    backgroundcolor=\color{lightgray!10},
    showspaces=false,
    showstringspaces=false,
    breaklines=true,
    breakatwhitespace=true,
    tabsize=4
}

\title{Design for light-based spherical aberration correction of ultrafast electron microscopes}

\author{Marius Constantin Chirita Mihaila,\authormark{1,*} Martin Kozák,\authormark{1} }

\address{\authormark{1}Charles University, Faculty of Mathematics and Physics, Ke Karlovu 3, 121 16 Prague 2\\}

\email{\authormark{*}marius.chirita@matfyz.cuni.cz} 


\begin{abstract*} 
We theoretically demonstrate that ponderomotive interactions near the electron cross-over can be used for aberration correction in ultrafast electron microscopes. Highly magnified electron shadow images from Si$_3$N$_4$ thin films are utilized to visualize the distortions induced by spherical aberrations. Our simulations of electron-light interactions indicate that spherical aberrations can be compensated resulting in an aberration free angle of \SI{8.1}{mrad}. For achieving the necessary light distribution, we use a gradient descent algorithm to optimize Zernike polynomials and shape the light beam into a modified Gaussian and Laguerre-Gaussian beam. 

\end{abstract*}

\section{Introduction}
Ultrafast transmission electron microscopes (UTEMs) enable unprecedented temporal resolution in nanoscale imaging \cite{zewail2010four, arbouet2018ultrafast, montgomery_leonhardt_roehling_2021,morimoto2023attosecond}. Further advances are expected from the combination of highly coherent field emission sources \cite{feist2017ultrafast, houdellier2018development,zhu2020development} with aberration-corrected probes~\cite{feist2017ultrafast, arbouet2018ultrafast}. 

Aberration correction can be realized using static multipoles \cite{haider1998spherical,muller2006advancing, rose2009historical}, which has led to sub-Angström imaging in materials science~\cite{krivanek1999towards}. However, such correctors are limited to the correction of low-order aberrations due to the structure of the multipoles. Alternatively, material phase plates have been employed for spherical aberration correction~\cite{shiloh2018spherical,roitman2021shaping}, but they encounter issues related to electron losses and inelastic scattering.

Programmable and adaptive optics, such as spatial light modulators, have significantly advanced light optics, leading to important applications in astronomy, microscopy, and tomography~\cite{maurer2011spatial, Rubinsztein_Dunlop_2016, hampson2021adaptive}. In contrast, the development of programmable phase plates for electron optics is still in its infancy. However, promising innovations are emerging, including electrostatic Einzel lens arrays~\cite{VERBEECK201858, ibanez2022, ribet2023design, yu2023quantum} and miniaturized multipolar lenses~\cite{grillo2014}. In addition to these advancements, the interaction of electrons with optical near-fields has enabled coherent manipulation of electron waves~\cite{barwick2009photon,vanacore2018attosecond,vanacore2019ultrafast,konevcna2020electron,ben2021shaping,henke2021integrated,dahan2021imprinting,feist2022cavity,madan2022ultrafast,tsesses2023tunable,garcia2023spatiotemporal}.

 Free space interactions \cite{Bucksbaum1988,freimund2001observation,dwyer2006femtosecond, Kozak2018, schwartz2019laser, axelrod2020observation, chirita2022transverse, tsarev2023nonlinear, ebel2023inelastic, streshkova2024electron, streshkova2024monochromatization,velasco2024free} explore the potential of pure optical fields influencing electron wave function, opening avenues for new experimental approaches. Recent studies have expanded on this by investigating possibilities for correcting spherical aberrations in electron lenses through controlled light interactions \cite{de2021optical, uesugi2021electron, uesugi2022properties}.

The spatial resolution of Ultrafast Transmission Electron Microscopes (UTEMs) can be constrained by the low probe current, which typically ranges from 1 to 100 fA~\cite{feist2017ultrafast}. This probe current is limited by the Boersch effects near the electron gun, where electron velocities are low, as well as by the use of small objective lens apertures. Introducing aberration correction techniques could enable the use of larger probe apertures, thereby increasing probe currents. This enhancement would reduce acquisition times while preserving or improving imaging resolution. 

Several diagnostic techniques have been developed to assess aberrations in electron lenses, including the analysis of diffractograms~\cite{saxton2000new}, and the use of electron Ronchigrams from various sample types, both amorphous and crystalline~\cite{cowley1979adjustment, lin1986calibration, krivanek1999towards, james1999practical, liu2005scanning, sawada2008measurement}. Additionally, the shadow imaging technique, which employs fine gratings to explore focal spot characteristics, provides valuable insights, particularly when Ronchigram methods are not feasible~\cite{spangenberg1942some, rempfer1985unipotential, rempfer1997simultaneous}.

 Here, we use ponderomotive electron beam shaping in free space near the electron cross-over. In contrast to the method in~\cite{uesugi2021electron,uesugi2022properties}, which employs a donut beam for ponderomotive correction and introduces additional defocus due to its quadratic shape near the optical axis, our approach employs Zernike polynomials to shape the laser beam into nearly ideal quartic and inverse quartic intensity distributions. Our approach avoids lensing effects which simplifies the design of the ponderomotive correction method when implemented in existing electron microscopes. Furthermore, to develop new aberration correction techniques, it is essential to accurately identify and characterize these aberrations. As such, we provide ways to visualize the electron aberrations in ultrafast electron microscopes by simulating highly magnified shadow images of Silicon Nitride (Si$_3$N$_4$) with patterned holes. The shadow images are acquired with the electron beam focused before the sample, and while negative spherical aberration phase is more common, we consider both negative and positive values. Our method could increase the aberration-free angle at high $C_s$ settings and improve the resolution of ultrafast electron microscopes. 
 
 To investigate the interaction of the aberrated electron beam with the laser field and sample, we employ both ray and wave optics simulations. The ray optics method offers a straightforward, intuitive understanding of the electron trajectories and their deviations due to lens aberrations and laser-induced ponderomotive forces. On the other hand, the wave optics approach provides a comprehensive analysis of the electron wavefronts, capturing detailed phase information and interference effects. By utilizing both methods, we aim to ensure that our theoretical predictions are accurate and in agreement with each other. 

\section{Theoretical Considerations}
\begin{figure}[t]
   \centering      
   \includegraphics[width=12.0cm]{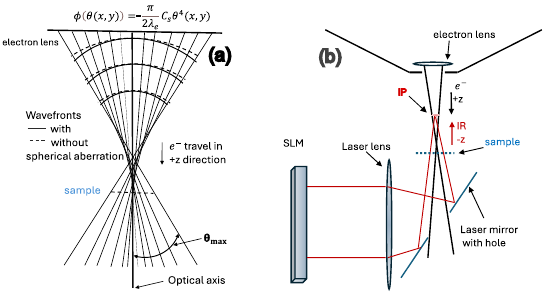}
   
 \caption{(a) Electron trajectories and wavefronts with spherical aberrations (not to scale). The phase shift distortion caused by spherical aberrations of the electron lens is described as a function of semi-convergence angle $\theta (x,y)$ with the spatial coordinates x and y defined at the aperture plane of the lens. The semi-convergence angle $\theta$ is defined as the angle between the optical axis and the outermost ray of the electron beam passing through the lens.
 (b) The experimental setup designed to investigate the ponderomotive spherical aberration correction of an electron lens. The interaction plane (IP) between the laser and electron pulse is in the focal point of the laser beam. The spatial light modulator (SLM) on the left shapes and modulates the infrared (IR) laser light ($E_L < \SI{1}{\micro J}$, pulse length $\tau_L \sim \SI{250}{fs}$, $\lambda_L = \SI{1030}{nm}$). The shaped IR light is then focused by a lens before it interacts with the electron beam. The laser mirror, which has a central hole, allows the electron beam to pass through while reflecting the IR light towards the IP. The electron beam (pulse length $\tau_e \sim 2 \tau_L$) starting at the pole piece of an ultrafast scanning electron microscope (USEM), passes through the IP, the sample, and the hole in the mirror. The electron cross-over is between the focal spot of the shaped light and sample. The electrons propagate further to the detection system (not shown here).}
 \label{fig:experimental setup}
\end{figure}

Spherical aberrations in an optical system, such as an electron lens, result in marginal rays (those farthest from the optical axis) focusing at a point before the image plane (see Fig.~\ref{fig:experimental setup}(a)). This phenomenon can significantly impact image clarity and focus. In describing the aberration function of an electron lens, we typically consider both primary spherical aberrations and defocus:
\begin{equation}
    \phi (\theta)= \frac{\pi}{2\lambda_e}( C_s\theta^4-2\Delta z\theta^2),
    \label{eq: Aberration function wave}
\end{equation}
where $\theta = \theta(x,y)$ is the semi-convergence angle, $\lambda_e = h/(\gamma m_e v)$ is the electron wavelength with $v$ the relativistic electron velocity, $m_e$ the electron mass, $\gamma = 1/\sqrt{1-v^2/c^2}$ the relativistic factor, $C_s$ is the spherical aberration coefficient and $\Delta z$ is the defocus that can be tuned. Spherical aberrations can be effectively minimized by adjusting the defocus to the Scherzer defocus, specifically by setting $\Delta z = 0.5C_s\theta_{\text{max}}^2$~\cite{scherzer1949theoretical}, where $\theta_{\text{max}}$ is the maximum convergence semi-angle.

The use of ponderomotive interaction to correct spherical aberrations in electron lenses illustrates an interplay between the manipulation of light and electron behavior. By fine-tuning the light field’s intensity and spatial configuration, we can precisely adjust electron wavefronts. In the non-recoil approximation, where the interaction with light induces only minor changes to the electron's energy compared to its mean energy it is demonstrated in~\cite{PhysRevLett.126.123901} that the Dirac equation simplifies to an effective Schrödinger equation:
\begin{equation}\label{eq:effSE}
    (\partial_t +\bm{v}\cdot \nabla)\psi_{\perp}(\bm{r},t) = -\frac{i}{\hbar} U (\bm{r},t) \psi_{\perp}(\bm{r},t).
\end{equation}
The ponderomotive pontential reads:
\begin{eqnarray}
    U(\bm r,t) &=& \frac{e^2}{2m_e \gamma} \left[ A_x^2 (\bm r, t) + A_y^2 (\bm r, t) + \frac{1}{\gamma^2} A_z^2 (\bm r, t) \right],
    \label{eq:intPotential}
\end{eqnarray}
where $\bm{A}$ is the vector potential. Upon integrating Eq.~\eqref{eq:effSE}, the evolution of the electron state $\psi_{\perp}$ after ($t_0$) and before ($t_1$) the interaction with light takes the following form:
\begin{eqnarray} \label{eq:scatteringTrafo}
    \psi_{\perp}(\bm{r}+\bm v t_1,t_1) &=& \psi_{\perp}(\bm{r}+\bm{v}t_0,t_0) \times  \exp \left[-\frac{i}{\hbar} \int_{t_0}^{t_1} \diff t' \, U(\bm{r}+ \bm{v}t', t') \right]. \nonumber 
\end{eqnarray}
Following on~\cite{chirita2022transverse}, the phase shift acquired by the electron during interaction with a counter propagating laser beam takes the following analytical form:
\begin{equation}
\begin{aligned}
    \varphi(x,y) &= \left[-\frac{1}{\hbar} \int_{t_0}^{t_1} \mathrm{d} t' \, U(\bm{r} + \bm{v}t', t') \right] \approx -\frac{\alpha}{2\pi (1+ \beta)} \frac{E_L \lambda_L^2}{E_e} \frac{ g^2(x,y)}{\int \mathrm{d} x \mathrm{d} y\, g^2(x,y)}.
\end{aligned}
\label{eq:phase_shift}
\end{equation}
Here, $\alpha$ is the fine structure constant, $\beta = v/c$ is the velocity of the electron in units of the speed of light, $\lambda_L$ is the laser wavelength, $E_e = \gamma m_e c^2$ the relativistic electron energy, $E_L$ the laser pulse energy, and $g^2(x,y)$ represents the spatial intensity profile of the laser. It should be noted that the calculation of the phase shift (Eq.~\ref{eq:phase_shift}) does not consider spontaneous processes, electron spin, and uses paraxial approximation and monochromaticity for both the electrons and light.

To correct for the electron aberrations acquired in the magnetostatic lens, we need to introduce an opposite phase shift that cancels out the distortions. To achieve ideal aberration correction, the condition $\phi\bigl(\theta(x,y)\bigr) = -\varphi(x,y)$ must be satisfied, which implies that $g^2(x,y)$ should exhibit an inverse quartic or quartic profile depending on the sign of the spherical aberration phase of the electron lens.

\subsection{Ray optics}

The ray tracing model starts with the aberrated initial electrons state in the interaction plane (IP) with the laser beam:
\begin{equation}\label{eqn:initial_ray_state}
    \eta_1 = \begin{bmatrix} x \\ y \\ \theta_{0}(x) + \delta\theta_{a}(x) + \delta\theta_{L}(x)\\ \theta_{0}(y) + \delta\theta_{a}(y)+\delta\theta_{L}(y) \end{bmatrix},
\end{equation}
where $x, y$ are coordinates distributed within a circular beam of $r_0$, \(\theta_{0}(x) = -\frac{x}{d_c}\), and \(\theta_{0}(y) = -\frac{y}{d_c}\) are the initial propagation angles with $d_c$ being the distance from the IP to the geometrical focal spot (cross-over) of the electrons. Adjustments to the initial angles due to the electron lens aberrations and the laser-induced ponderomotive forces are given as:
\begin{equation}
    \delta\theta_{a}(x) = \frac{1}{k} \frac{\partial \phi\bigl(\theta(x,y)\bigr)}{\partial x}, \quad \delta\theta_{a}(y) = \frac{1}{k} \frac{\partial \phi\bigl(\theta(x,y)\bigr)}{\partial y},
\end{equation}
\begin{equation}
     \delta\theta_{L}(x) = \frac{1}{k} \frac{\partial \varphi(x,y)}{\partial x}, \quad  \delta\theta_{L}(y) = \frac{1}{k} \frac{\partial \varphi(x,y)}{\partial y},
\end{equation}
where  \(k = \frac{2\pi}{\lambda_e}\) is the electron wave number.

The rays are propagated to the sample plane (below the electron cross-over):

\begin{equation}
    \eta_2 = S_1 \eta_1,
\end{equation}
using the free space propagation matrix $S_i$:

\begin{equation}
S_i = 
    \begin{bmatrix}
        1 & 0 & d_i & 0 \\
        0 & 1 & 0 & d_i \\
        0 & 0 & 1 & 0 \\
        0 & 0 & 0 & 1 \\
    \end{bmatrix},
\end{equation}
where $d_1 = d_c + s$ is the distance from the IP to the sample plane, and $s$ is the distance from the electron cross-over to the sample.

The transmission function of the sample, $T_s$, is defined by a hexagonal grid of circles with transmission 1, each with a radius $R$:
\begin{equation}
    T_s(x_2, y_2) = \left\{
    \begin{array}{ll}
        1 & \text{if within hexagonal aperture} \\
        0 & \text{otherwise}
    \end{array}
    \right.
\end{equation}

The interaction is modeled as:
\begin{equation}
    \eta_{\text{t}} = \eta_2 \cdot T_s(x_2, y_2),
\end{equation}
where \( \eta_{\text{t}} \) contains only those electrons that have passed through the apertures. The electrons that pass through \(T_s\) are further propagated to the detector over a distance $d_2$ using:
\begin{equation}
    \eta_3 = S_2 \eta_{\text{t}},
\end{equation}

The final imaging on the detector is analyzed by mapping the positions and densities from the propagated electron rays, allowing a detailed reconstruction of the aberrated electron beam's interaction with the laser beam and sample.

\subsection{Wave optics}

We use the Fresnel Propagation (FP) method to simulate the the electron beam from its initial interaction with the laser beam to the point of contact with the sample. This choice is favored over the Fast Fourier Transform (FFT) because it allows for detailed visualization of the field shape beneath the electron focal point. Subsequently, we apply an FFT to model the electron beam's final projection onto the detector.

The initial electron state reads:

\begin{equation}
    \psi_0(x,y)= T_a\cdot T_l \cdot e^{i(\phi(\theta(x,y))+\varphi(x,y))},
    \label{eqn: initial electron state}
\end{equation}
where $T_a$ and $T_l$ are the aperture and lens transmission functions, respectively (see Appendix~\ref{sec:A} for details).

The initial wave function $\psi_0(x,y)$ is then propagated for $d_1$ at the sample plane and can be expressed as:
\begin{equation}
    \psi_p(x_2,y_2)= \text{FP}(\psi_0(x,y)),
    \label{eqn: Field_probe 1}
\end{equation}
where $x_2$ and $y_2$ are the real space units in the Fourier space. In the Fourier plane, the electron wave function is multiplied by the transmission function of the sample $T_{s}$ and propagated using FFT for $d_2$ to the detector. The intensity at the detector plane $I_{d}$ reads as:
\begin{equation}
    I_{d} = \left| \text{FFT} \left( \psi_p(x_2,y_2)\cdot T_{s} \right) \right|^2.
    \label{eq: Intensity detector}
\end{equation}

\section{\label{sec:level4}Electron beam simulations with shaped light}
This section describes a possible experimental design for testing ponderomotive aberration correction in an USEM operated in transmission mode, as illustrated in Fig.~\ref{fig:experimental setup}. Before reaching the IP, the infrared (IR) laser beam passes through a mirror with a central hole of radius $r_h = \SI{125}{\micro\metre}$. Notably, this small hole has a negligible impact on the light field distribution, as demonstrated in recent studies~\cite{ChiritaMihaila2022}. The hole's size is significantly smaller—6 times—than those used in previous configurations~\cite{chirita2022transverse}, ensuring minimal disturbance to the laser beam.

\begin{figure}[t]
   \centering      
   \includegraphics[width=12.0cm]{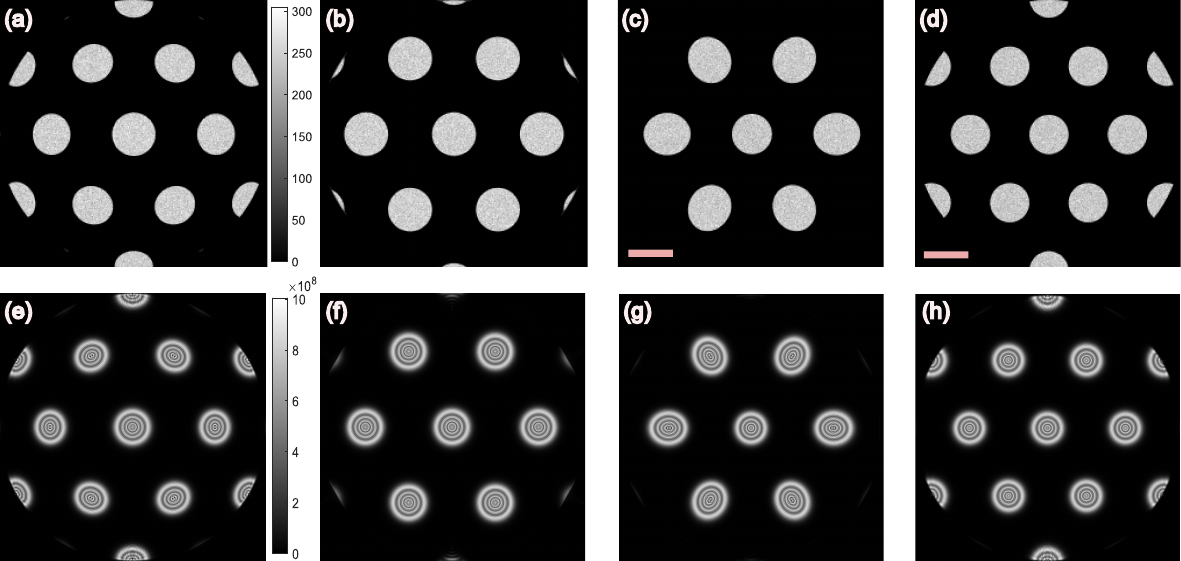}
 \caption{Comparative illustration of electron-light interaction models at the detector plane ($E_0 = \SI{30}{keV}$, $\theta_{\text{max}} = \SI{8.1}{mrad}$). Deviation from the round shape in the holes indicate primary spherical aberrations. Top row depicts the ray optics approach: (a)  Electron beam distribution with the laser off and negative aberration phase of the electron lens. (b) Laser on (inverse quartic laser beam distribution) at \SI{0.2}{\micro J} with $\delta\theta_{a}(x,y) = - \delta\theta_{L}(x,y)$ and negative aberration phase. (c)  Electron beam distribution with the laser off (scale bar \SI{0.532}{mm}) and positive aberration phase. (d) Laser on (quartic laser beam distribution) at \SI{0.2}{\micro J} with $\delta\theta_{a}(x,y) = - \delta\theta_{L}(x,y)$ and positive aberration phase (scale bar \SI{0.535}{mm}). Bottom row presents the wave optics model: (e) Electron beam distribution with the laser off and negative aberration phase. (f) Laser on at \SI{0.2}{\micro J} with $\phi\bigl(\theta(x,y)\bigr) = -\varphi(x,y)$ and negative aberration phase. (g) Electron distribution at the detector with laser off and positive aberration phase. (h) Laser on at \SI{0.2}{\micro J} with $\phi\bigl(\theta(x,y)\bigr) = -\varphi(x,y)$ and positive aberration phase. The structure of the original sample can be seen in Fig.~\ref{fig:Original sample} in Appendix~\ref{sec:A}. To be noted that most electron lenses have a negative spherical aberration phase as depicted in Fig~\ref{fig:experimental setup} (a).}
 \label{fig:Ray vs Wave}
\end{figure}

Based on the simulation analysis described in the previous chapter, the following parameters have been established and are used for the simulations in Fig.~\ref{fig:Ray vs Wave}, as follows: The electron mean energy is set to $E_0 = \SI{30}{keV}$. The electron lens has a focal length $f = \SI{8}{mm}$ ($C_s \propto  f^3\sim\SI{80}{mm}$, $\Delta z=0$), with a beam diameter at the lens of $\SI{130}{\micro m}$ leading to $\theta_{\text{max}} = \SI{8.1}{mrad}$. To be noted that the simulations from Fig.~\ref{fig:Ray vs Wave} start in the IP. At the IP, the electron beam radius is $\rho_{\text{IP}} = \SI{5}{\micro m}$. This beam radius reduces to $\rho_{\text{sample}}\sim\SI{0.325}{\micro\metre}$ at the sample plane after propagation for $d_1 = d_c + s = \SI{0.662}{mm}$ (where $d_c = \SI{0.62}{mm}$ and $s = \SI{0.042}{mm}$). Furthermore, at the mirror hole, which is positioned $\SI{12}{mm}$ below the sample, the electron beam radius expands to $\rho_{\text{mirror}}=\SI{98}{\micro\metre}<r_h$. Additionaly, the distance travelled by the electron from the sample to the position-sensitive detector is $d_2 \sim \SI{200}{mm}$, fitting within existing USEM chambers without requiring modifications.

The simulated sample used is Si$_3$N$_4$ thin film with patterned holes ($R = \SI{50}{nm}$, pitch \SI{200}{nm}) because it has almost ideal transmission at the laser wavelength $\lambda_L = \SI{1030}{nm}$ and film thickness of $\sim\SI{260}{nm}$,  does not get damaged under defocused laser illumination, allows for high magnification imaging, and is commercially available. The thickness of Si$_3$N$_4$  would typically prevent most electrons from passing through. As such, we neglect the thickness of the film in our simulations and assume either 0 or 1 transmission. While the electron pulse duration is longer than the laser pulse, any potential interaction between the electrons and the laser field at the sample plane can be ruled out. At a distance $d_1$ from the interaction plane, any plasmonic excitations generated at the sample would have already decayed before the electrons reach the mask. Additionally, the defocused laser beam at the sample plane prevents the possibility of four-photon excitations, eliminating the formation of a Coulomb cloud at an intensity of $ \SI{7}{GW/cm^2}$.

Figure~\ref{fig:Ray vs Wave}(top) shows the ray and Fig.~\ref{fig:Ray vs Wave}(bottom) the wave optics simulations of the electron beam at the detector plane. Figures~\ref{fig:Ray vs Wave}(a),(e) and (c),(g) are the aberrated electron shadow images of Si$_3$N$_4$ when the spherical aberration phase of the electron lens is negative and positive, respectively. The electron distributions in Fig.~\ref{fig:Ray vs Wave}(b),(f) and Fig~\ref{fig:Ray vs Wave}(d),(h) are simulated with the additional interaction of the inverse quartic and quartic light intensity fields, respectively, at the optimal correction setting.  The intesities along the center of Fig.~\ref{fig:Ray vs Wave}(c),(d) are plotted in Fig.~\ref{fig:Intesity profiles along the line} in Appendix~\ref{sec:B} to illustrate the eliptic shape induced by aberrations. Note that our calculations assumed a fully coherent beam and neglected vibrations, higher-order aberrations, and other non-idealities such as Johnson noise~\cite{uhlemann2013thermal}.

\begin{figure}[t]
  \centering      
  \includegraphics[width=12cm]{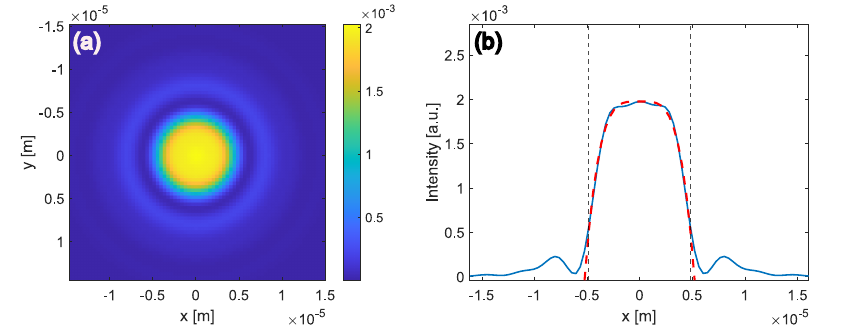}
  \caption{(a) The intensity distribution of an inverse quartic beam in focus with an NA of $\sim$ 0.16. (b) Extracted region of interest from (a) and $1 - (x/a)^4$ fit within the region $r_0\leq\SI{5}{\micro m}$ where the shape is inverse quartic. The vertical lines indicate where approximated inverse quartic shape ends. Furthermore, the coefficient of determination, $Q^2$, is calculated to evaluate the fit ($1 - (x/a)^4$) of the model. It is determined using the formula $Q^2 = 1 - \frac{\text{SSE}}{\text{SST}} = 0.983$, where SSE is the sum of squared errors between the observed and predicted values, and SST is the total sum of squares representing the total variance in the data. This shape can be used to correct for the negative spherical aberration phase of the electron wave as illustrated in Fig.~\ref{fig:experimental setup} (a).
  }
  \label{fig:inverse_quartic}
\end{figure}
  To enhance the efficiency of the ponderomotive aberration correction, we employed a custom phase to shape a Gaussian beam, optimized using a gradient descent algorithm based on a method similar to that described in Ref.~\cite{xie2015phase,gerchberg1972practical}. The efficiency $ \eta_{\text{eff}}$ is then defined as the ratio of the intensity within the aperture ($r \leq r_0$, $I_{\text{roi}}$) to the total intensity ($I_{\text{tot}}$).

\begin{figure}[h]
   \centering      
   \includegraphics[width=12cm]{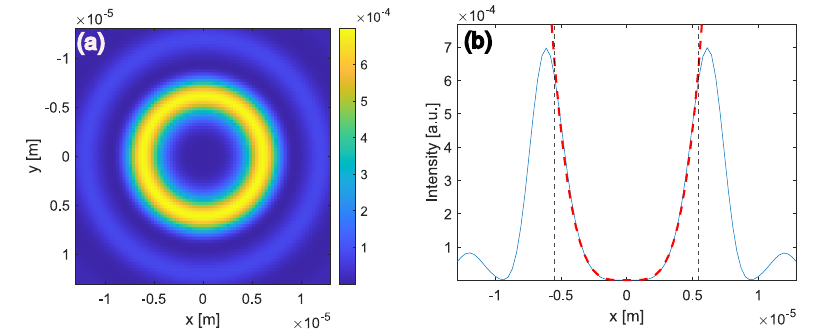}
   
 \caption{Laser beam shaping:
    (a) Shaped Gaussian beam into a modified LG beam using the phase shown in Fig.~\ref{fig:quartic phase shift}(c) from Appendix~\ref{sec:B}. (b) Fit along a line in the center of (a) showing nearly ideal quartic dependence, illustrating the required shape for ponderomotive spherical aberration correction of electron lenses. The numerical aperture of the lens of the simulated image (a) is $\text{NA} \sim 0.2$ and $Q^2 = 0.995$ for the fit function $ax^4$. This shape can be used to correct for the positive spherical aberration phase of the electron wave.
    }
 \label{fig:quartic_intensty}
\end{figure}

The phasemask $\phi_{\mathrm{iquartic}}$ applied to the initial Gaussian beam to shape it into an inverse quartic distribution (see Fig.~\ref{fig:inverse_quartic}) can be seen in Fig.~\ref{fig:quartic phase shift} (d).
The phasemask applied to the initial Gaussian beam to shape it into a quartic distribution $\phi_{\mathrm{quartic}}=\mathrm{modulo2\pi}[\phi_{\mathrm{LG03}}+\phi_{\mathrm{Zsum}}]$ is a sum of vortex phase plate of charge 3 and gradient descent optimized~\cite{nocedal1999numerical} Zernike phase masks (see Appendix~\ref{sec:B} for details). The results of these laser beam shaping can be visualized in Figs.~\ref{fig:inverse_quartic},\ref{fig:quartic_intensty} with $\eta_{\text{eff}} \sim 71\%$ for the inverse quartic beam and $\eta_{\text{eff}} \sim 27\%$ for the quartic beam. A numerical aperture of $\text{NA} = D_L/(2f_L) \sim 0.16$ has been used for the simulation in Fig.~\ref{fig:inverse_quartic} where $D_L$ is the light beam diameter at the lens with $f_L = \SI{25}{mm}$ the focal length. A numerical aperture of $\text{NA} \sim 0.2$ has been used for the simulation in Fig.~\ref{fig:quartic_intensty}.

To simulate the images in Fig.~\ref{fig:Ray vs Wave} (b), (f) and Fig.~\ref{fig:Ray vs Wave} (d), (h), the $g^2(x,y)$ in Eq.~\ref{eq:phase_shift} was set to $1 - r^4$ and $r^4$, respectively. For optimal spherical aberration correction, an initial pulse energy, $E_L$, of \SI{0.2}{\micro\joule} is necessary, as illustrated in Fig.~\ref{fig:Ray vs Wave}(b), (d), (f), (h). To account for the pulse energy distributed outside the region of interest, the total required energy, $E_{\text{tot}}$, is calculated as $E_L$ divided by the efficiency $\eta_{\text{eff}}$. Consequently,  $E_{\text{tot}}$ is determined to be $\sim$\SI{0.3}{\micro\joule} for the inverse quartic and $\sim$\SI{0.74}{\micro\joule} for the quartic distribution.

\section{Discussion and Conclusion}
In this study, we have theoretically demonstrated the correction of primary spherical aberration in lenses for ultrafast electron microscopes through ponderomotive interactions near the electron cross-over. To visualize distortions induced by spherical aberrations, we utilized highly magnified electron shadow images from Si$_3$N$_4$ thin films. The Gaussian laser beam was shaped using a combination of Zernike phase patterns to achieve the inverse quartic profile, and a combination of a vortex phase mask with Zernike phase patterns to produce the quartic profile. Additionally, probe size calculations before and after correction, considering the high Cs value, suggest a potential increase in resolution (see Appendix~\ref{sec:C} for details). Our simulations indicate that spherical aberrations could be corrected without introducing defocus, offering a potential advantage over methods that rely on an LG01 beam~\cite{uesugi2021electron,uesugi2022properties}. Furthermore, the light shape may be programmed to correct also for higher order aberrations. Nevertheless, the setup described here will allow for comparative testing to determine the most effective method. The experimental setup inside the vacuum is simple, involving one lens, one mirror with a hole in the center and one high resolution detection system. 

Despite the promising results, several limitations must be addressed. Drifts in the laser beam, which can lead to misalignments, represent the most significant issue in maintaining the precision of the beam shaping process (see Appendix~\ref{sec:C} for detailed analysis). Such drifts can cause spatial misalignments between the laser beam and the electron beam, impacting the effectiveness of the aberration correction and thereby degrading the probe shape and resolution. To minimize distortion, we recommend using a laser with pulse energy stability of less than $1\%$ rms. Additionally, employing an active Piezo beam pointing stabilization system with a position accuracy of approximately 4 nrad will further enhance stability. Also, placing a compact fiber laser directly on the electron microscope table would be beneficial. Moreover, operating the microscope at Scherzer defocus and measure an electron Ronchigram to show correction can reduce the required pulse energy. However, electron Ronchigrams are challenging to obtain using ultrafast electron microscopes. To the best of our knowledge, there are no peer-reviewed studies that have reported the acquisition of electron Ronchigrams in an ultrafast electron microscope. We believe that the primary reason for this is the low electron current inherent in ultrafast electron microscopes, which significantly degrades the interference pattern and makes Ronchigram acquisition impractical. Consequently, we identified shadowgraphy as a more suitable and effective method for aberration diagnostics in this context.

Increasing $\theta_{\text{max}}$ requires more $E_L$. Also, we have presented a model utilizing a moderate NA up to $\sim$ 0.2. Note that a NA$>$0.2 might impose additional constraints on the calculation of the ponderomotive phase shift, as the electrons may traverse regions with intensity gradients during the interaction. 
The moderate NA leads to an extended depth of focus, which is larger than both the electron propagation distance during the interaction and the axial spread of the electron wave packet. In our configuration, each electron experiences the full laser pulse within the focal region with an interaction length of about \SI{20}{\micro m}. If an electron pulse shorter than the optical cycle is used, the scheme remains effective. However, if the duration of the electron pulse is too long, parts of the pulse may interact with the defocused regions of the optical field, which could degrade the overall quality of the interaction. By employing a shaping radius $r_0>\SI{5}{\micro m}$, one can use a smaller NA and reduce relative vibration issues, albeit at the cost of increased $E_L$. Furthermore, for a NA of 0.2, 1 µJ pulse energy and 250 fs pulse length, the intensity at the sample position ($\SI{0.66}{mm}$ away from the focus) would be $\sim\SI{7}{GW/cm^2}$, which is well below the damage threshold of Si$_3$N$_4$ of $\sim\SI{2000}{GW/cm^2}$~\cite{tan2019silicon}.

The resolution of the detector also imposes constraints on the minimum $\delta \theta_L$ that can be observed. For instance, a detector resolution of  $\delta x = \SI{14}{\micro m}$ at a distance  $d_2 \sim \SI{20}{cm}$ requires $\delta\theta_L \sim \delta x/d_2 = \SI{70}{\micro rad}$ deflections. This constraint highlights the need for high-resolution detectors, or high magnification systems to fully characterise the benefits of ponderomotive aberration correction. Moreover, previous research has demonstrated that phase-only correction based on materials~\cite{roitman2021shaping} is feasible even with an electron beam diameter of \SI{150}{\micro\metre} at the objective lens aperture. Furthermore, using larger apertures with implementation of ponderomotive aberration correction, together with temporal pulse compression techniques to manage high number of electrons/pulse~\cite{van2010compression,zhou2017femtosecond,williams2017active} could improve even more the electron signal. Lastly, the IP can be placed near the crossover point of the condenser lens of a microscope equipped with a Schottky or cold field emission gun, both of which offer excellent transverse coherence~\cite{feist2017ultrafast,houdellier2018development}.

In conclusion, we present a realistic method for spherical aberration correction in ultrafast electron microscopy, utilizing the interaction between shaped light fields and electron beams. The approach described here has the potential to enhance the resolution of UTEMs, ultrafast Lorentz microscopes~\cite{berruto2018laser,rubiano2018nanoscale}, single-shot ultrafast electron microscopes~\cite{kim2008imaging}, ultrafast cryo-TEMs~\cite{du2024development} and others.

\appendix

\section{\label{sec:A} Electron beam simulations}

\subsection*{Fresnel Propagator}
The Fresnel Propagator is computed using MATLAB~\cite{MATLAB} and models the propagation of an electron beam in a near-field setup under the paraxial approximation, which is crucial for understanding the wavefront evolution over short distances $d_1$. Due to the large grid dimension required to fulfil the sampling condition, the simulation starts at the IP.

The Fresnel Propagator ($\text{FP}(\psi_0(x,y))$) is computed through several sequential steps involving the transformation of spatial coordinates into frequency coordinates and the application of a quadratic phase factor~\cite{voelz2011computational}. Initially, the frequency coordinates \( f_x \) and \( f_y \) are calculated based on the grid dimensions \( N_x \times N_y  = 23500 \times 23500 \) and the respective grid spacings \( dx \) and \( dy \). Critical sampling has been used, which reads $dx = \sqrt{\lambda_e  d_c / N_x};
$ to ensure that there are no artifacts in the simulation. These coordinates are defined as:
\begin{equation}
f_x = \frac{-N_x/2 \text{ to } N_x/2-1}{N_x \cdot dx}, \quad f_y = \frac{-N_y/2 \text{ to } N_y/2-1}{N_y \cdot dy}
\end{equation}

A meshgrid of these frequencies, \( F_X \) and \( F_Y \), is then constructed to represent the spatial frequency components across the computational grid.

The transfer function \( H \) of the Fresnel propagator is computed next. This function incorporates the wavelength \( \lambda_e \) of the electrons and the propagation distance \( d_1 \), and is given by:
\begin{equation}
H = \exp\left(-i \pi \lambda_e d_1 (F_X^2 + F_Y^2)\right)
\end{equation}
This function models the phase alterations due to diffraction in free space under the Fresnel approximation.

The application of the Fresnel propagator to an input field $u_1 = \psi_0(x,y)$ involves a series of Fourier transformations. The input field is first centered and transformed to the frequency domain:
\begin{equation}
U_1 = \text{FFT}(\text{fftshift}(u_1))
\end{equation}
It is then multiplied by the transfer function \( H \) to simulate propagation:
\begin{equation}
U_2 = H \cdot U_1
\end{equation}
Finally, the propagated wave function is reconstructed in the spatial domain by reversing the previous Fourier transform and centering the result:
\begin{equation}
u_2 = \text{ifftshift}(\text{IFFT}(U_2)),
\end{equation}
where $u_2 = \psi_p(x_2,y_2)$ in our simulations.

\subsection*{Transmission function of the aperture, lens and sample }

The aperture transmission function, \( T_a \), models the physical constraints imposed by the electron microscope's aperture. It is defined as a circular mask:
\begin{equation}
T_{a}(x, y) = 
\begin{cases} 
1 & \text{if } \sqrt{x^2 + y^2} \leq r_{\text{a}} \\
0 & \text{otherwise}
\end{cases}
\end{equation}
where \( r_{\text{a}} = 5 \times 10^{-6} \) meters is the radius of the aperture.

The lens transmission function incorporates the effects of focusing:
\begin{equation}
T_{l}(x, y) = \exp\left(-i \cdot \frac{k}{2d_c} (x^2 + y^2)\right)
\end{equation}

\begin{figure}[t]
   \centering      
   \includegraphics[width=12.0cm]{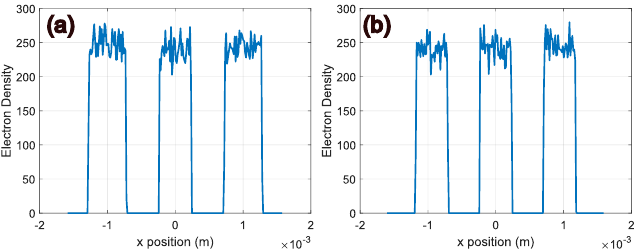}
   
 \caption{Coresponding intesity profile along a line around the center of images from Fig~\ref{fig:Ray vs Wave}(c),(d) showing the position dependent asymmetry induced by aberrations.}
 \label{fig:Intesity profiles along the line}
\end{figure}
The sample grid is constructed with a horizontal pitch \(P_{x2} = 4R\) and a vertical pitch \(P_{y2} = 2\sqrt{3}R\). Circles in even rows are aligned, while those in odd rows are shifted horizontally by \(P_{x2}/2\). The transmission function can be expressed as:

\begin{equation}
    T_s(x_2, y_2) = 
    \begin{cases} 
    1 & \text{if } (x_2 - m'P_{x2} - \delta_m')^2 + (y_2 - n'P_{y2})^2 \leq R^2 \\
    0 & \text{otherwise}
    \end{cases}
\end{equation}

where $m'$ and $n'$ are integers representing the grid indices, and $\delta_m'$ is $0$ for even $n'$ and $P_{x2}/2$ for odd $n'$.

\begin{figure}[h]
   \centering      
   \includegraphics[width=6cm]{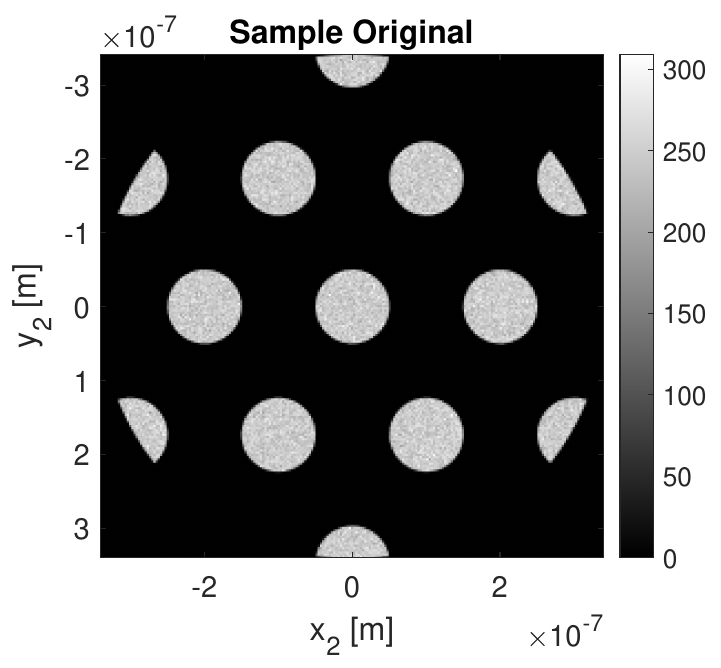}
   
 \caption{The Si$_3$N$_4$ thin film at the sample plane. The field of view has a radius of $\SI{0.325}{\micro m}$.}
 \label{fig:Original sample}
\end{figure}

\section{\label{sec:B}Laser beam shaping}

\subsection*{Zernike Polynomials}

Zernike polynomials are a set of orthogonal polynomials that are defined on the unit disk. They are especially important in systems such as optical systems and wavefront analysis where radial and azimuthal symmetry are key. In this section, we first present the mathematical formulation of these polynomials and then explain how we use them to shape the phase of a light beam. The Zernike polynomials~\cite{lakshminarayanan2011zernike} $Z_n^m(r, \theta_z)$ are defined in terms of radial and azimuthal coordinates as follows:
\begin{equation}
Z_n^m(r, \theta_z) = R_n^m(r) \cdot e^{i m \theta_z}
\end{equation}
 where $r = \sqrt{x^2 + y^2}$ is the radial coordinate ranging from 0 to 1, and $\theta_z$ is the azimuthal angle. Here, $ n $ is a non-negative integer representing the radial degree, and $m$ is an integer representing the azimuthal frequency such that $n - |m|$ is even and $|m| \leq n$. The angle $\theta_z$ (expressed in Cartesian coordinates) is defined as follows resulting in values from $-\pi$ to $\pi$:

\begin{equation}
\theta_z = \begin{cases} 
\tan^{-1}\left(\frac{y}{x}\right) & \text{if } x > 0, \\
\tan^{-1}\left(\frac{y}{x}\right) + \pi & \text{if } x < 0 \text{ and } y \geq 0, \\
\tan^{-1}\left(\frac{y}{x}\right) - \pi & \text{if } x < 0 \text{ and } y < 0, \\
\frac{\pi}{2} & \text{if } x = 0 \text{ and } y > 0, \\
-\frac{\pi}{2} & \text{if } x = 0 \text{ and } y < 0, \\
0 & \text{if } x = 0 \text{ and } y = 0.
\end{cases}
\end{equation}

\begin{figure}[t]
  \centering      
   \includegraphics[width=12.0cm]{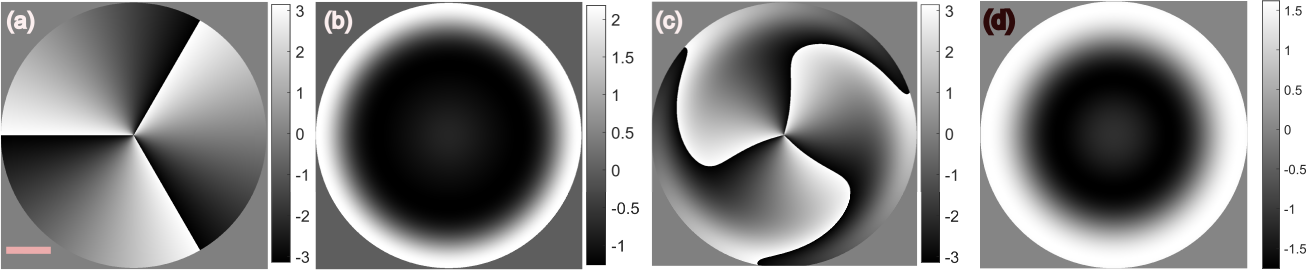}
   
 \caption{(a) The Vortex phase mask $\phi_{\mathrm{LG03}}$ used as an initial phase (scale bar $\SI{1.7}{mm}$) in the gradient descent optimisation algorithm and (b) the Zernike phase mask $\phi_{\mathrm{Zsum}}$ calculated using the gradient descent algorithm to shape the LG03 into a mode with quartic profile. (c) $\phi_{\mathrm{quartic}}=\mathrm{modulo2\pi}[\phi_{\mathrm{LG03}}+\phi_{\mathrm{Zsum}}]$. (d) $\phi_{\mathrm{iquartic}}$ the phasemask used to shape the source Gaussian into having an inverse quartic shape.}
 \label{fig:quartic phase shift}
\end{figure}

The radial component $R_n^m(r)$ of Zernike polynomials is defined as:
\begin{equation}
R_n^m(r) = \sum_{k=0}^{\frac{n - |m|}{2}} (-1)^k \frac{(n-k)!}{k! \left( \frac{n + |m|}{2} - k \right)! \left( \frac{n - |m|}{2} - k \right)!} r^{n - 2k}
\end{equation}
Figure~\ref{fig:quartic phase shift}(a) shows the phase mask $\phi_{\text{LG03}} = \arg\left(\exp(i \cdot (3) \cdot \theta_z)\right)$ and Fig.~\ref{fig:quartic phase shift}(b) the optimized Zernike phase mask:
\[
\begin{aligned}
\phi_{\mathrm{Zsum}} &= a_5 \cdot \pi \cdot Z_{2}^0(r, \theta_z) + a_{13} \cdot \pi \cdot Z_{4}^0(r, \theta_z) \\
&\quad + a_{25} \cdot \pi \cdot Z_{6}^0(r, \theta_z) + a_{41} \cdot \pi \cdot Z_{8}^0(r, \theta_z) \\
&\quad + a_{61} \cdot \pi \cdot Z_{10}^0(r, \theta_z) + a_{85} \cdot \pi \cdot Z_{12}^0(r, \theta_z).
\end{aligned}
\]
The coefficients of the Zernike phase masks, $a_5$, $a_{13}$, $a_{25}$, $a_{41}$, $a_{61}$, and $a_{85}$, have been optimized using the gradient descent algorithm described below. The optimized coefficients for generating the inverse quartic beam:
\[
\begin{aligned}
[a_5, a_{13}, a_{25}, &a_{41}, a_{61}, a_{85}] = [0.6547, 0.3183, -0.1863\\
&, -0.1290, -0.0024, 0.0298].
\end{aligned}
\]
The optimized coefficients to generate the quartic beam read:
\[
\begin{aligned}
[a_5, a_{13}, a_{25}, &a_{41}, a_{61}, a_{85}] = [0.6937, 0.0148, -0.3022\\
&, 0.0739, 0.0354, -0.0212].
\end{aligned}
\]

\subsection*{Optimization algorithm}

A gradient descent algorithm is employed to shape the amplitude distribution of an initial Gaussian laser beam, aiming to align it closely with a predefined target profile showing an inverse and quartic spatial distribution.

The optimization process starts in the real space (SLM plane) with a Gaussian source field $A(x_r,y_r) = \sqrt{\exp{(-2r^2/w_{00}^2)}}$, with ($w_{00} = \SI{3.8}{mm}$). The computational grid is defined with $ N_{x_r} = 8000 $ and $ N_{y_r} = N_{x_r} $, representing the number of pixels along the $x_r$ and $y_r$ directions respectively. The physical side lengths in the $x_r$  and $y_r$ directions are \( L_1 = N_{x_r} \cdot d_{x_r} \) and \( L_2 = N_{y_r} \cdot d_{y_r} \), where \( d_{x_r} = d_{y_r} = 8 \times 10^{-6} \, \text{m} \) is chosen such that it corresponds to the pixel size of an SLM. The physical dimensions of $N_{\text{xslm}}\times N_{\text{yslm}} = 1920 \times 1080$ pixels, creates an SLM side length of \( L_{1{\text{slm}}}  = N_{\text{xslm}} \cdot d_{x_r} \) and \( L_{2{\text{slm}}} = N_{\text{yslm}} \cdot d_{y_r} \). An SLM mask $T_{\text{slm}}(x_r,y_r)$ is applied on $A(x_r,y_r)$ during each iteration and reads:
\begin{equation}
  T_{\text{slm}}(x_r,y_r) = 
  \begin{cases} 
    1 & \text{for } r_\text{m} \leq L_{2{\text{slm}}}/2 \\
    0 & \text{otherwise}
  \end{cases}
  \label{eqn:Masking Function SLM}
\end{equation}
where $r_{\text{m}}$ is the radius of the region of interest.

The initial LG03 phase ($\phi_{\mathrm{LG03}}$) which is constrained in real space during the optimisation, combined with additional Zernike phase patterns ($\phi_{\text{Zsum}}$), determines the profile of the reconstructed amplitude  $R(x_f,y_f)$ in the Fourier plane. The choice of LG03 phase was not arbitrary but was informed by testing different modes, specifically LG01, LG02, LG03, and LG04. Among these, LG03 yielded the most favorable results in terms of both convergence and the final optimized solution. To be noted that we employ phase-only methods to shape the light beam. However, incorporating both phase and amplitude modulation could potentially enhance the quality of the reconstructed intensity~\cite{monmayrant2004new}. Prior to comparing the reconstructed and target amplitudes, an amplitude constraint $T_f(x_f,y_f)$ is applied in the Fourier plane to the reconstructed amplitude. This normalization step ensures the optimization focuses on refining the beam's profile by maintaining the beam's energy within preset limits. 

\begin{equation}
  T_f (x_f,y_f) = 
  \begin{cases} 
    1 & \text{for } r_{\text{m}} < \SI{5.8}{\micro m} \\
    0 & \text{otherwise}
  \end{cases}
  \label{eqn:Masking Function}
\end{equation}

\begin{equation}
  R(x_f,y_f) = \text{FFT}\left(A(x_r,y_r) e^{i\phi_{\text{LG03}}} e^{i\phi_{\text{Zsum}}}\right)\cdot T_f,
  \label{eqn:Reconstructed amplitude}
\end{equation}
with $x_f$ and $y_f$ being the real space coordinates in the Fourier plane.

The target amplitude $T(x_f,y_f)$ features a quartic distribution, whose edges are smoothed using a convolution with a Gaussian $G(x_f,y_f)$ to facilitate smooth transitions and reduce potential discontinuities during the Fourier transformation of the initial source field:

\begin{equation}
T(x_f, y_f) = (r^4 \cdot T_f (x_f,y_f)) * G(x_f, y_f),\label{eqn:Target amplitude}
\end{equation}
where $*$ denotes the convolution operation.

The correlation coefficient $c_r$ between the reconstructed beam $R(x_f,y_f)$ and the target beam $T(x_f,y_f)$, which quantifies their similarity, is computed using the MATLAB function \texttt{corr2}. The correlation coefficient is given by the formula:

\begin{equation}
c_r = \frac{\sum_{m'} \sum_{n'} (R_{m'n'} - \overline{R})(T_{m'n'} - \overline{T})}{\sqrt{\left(\sum_{m'} \sum_{n'} (R_{m'n'} - \overline{R})^2\right) \left(\sum_{m'} \sum_{n'} (T_{m'n'} - \overline{T})^2\right)}},
\label{eqn:correlation_coefficient}
\end{equation}
where $\overline{R}$ and $\overline{T}$ represent the mean values of the reconstructed and target beams, respectively. These means are computed as:

\begin{equation}
\begin{aligned}
\overline{R} &= \frac{1}{N_{x_r}N_{y_r}} \sum_{m'=1}^{N_{x_r}} \sum_{n'=1}^{N_{y_r}} R_{m'n'}, \\
\overline{T} &= \frac{1}{N_{x_r}N_{y_r}} \sum_{m'=1}^{N_{x_r}} \sum_{n'=1}^{N_{y_r}} T_{m'n'}.
\end{aligned}
\end{equation}

The indices $m'$ and $n'$ represent the row and column indices. Here $R_{m'n'}$ and $T_{m'n'}$ represent the individual elements of $R(x_f,y_f)$ and $T(x_f,y_f)$ matrices, respectively. This coefficient measures the degree to which the two fields are linearly related to each other, aiding in optimizing the beam shaping process by quantifying the alignment between the actual output and the desired outcome.

The core of the optimization process for adjusting the Zernike polynomial coefficients in the phase mask on an SLM involves the gradient descent algorithm, which utilizes $c_r$ as the basis for the cost function. The objective is to maximize $c_r$ between the $R(x_f, y_f)$ and  $T(x_f, y_f)$, which is computed as follows:

\textbf{Cost Function:} The cost function \( C(\mathbf{a}) \) is defined as the negative of the correlation coefficient $c_r$ between the reconstructed beam and the target beam, intended to be minimized:
\begin{equation}
C(\mathbf{a}) = -\text{corr2}(R(x_f, y_f; \mathbf{a}), T(x_f, y_f)),
\end{equation}
where \( \mathbf{a} \) represents the vector of coefficients for the Zernike polynomials.

\textbf{Gradient Computation:} The gradient of the cost function \( \nabla C(\mathbf{a}) \) with respect to the coefficients vector \( \mathbf{a} \) is computed numerically:
\begin{equation}
\nabla C(\mathbf{a}) = \left[ \frac{\partial C}{\partial a_1}, \frac{\partial C}{\partial a_2}, \ldots, \frac{\partial C}{\partial a_k} \right]^T,
\end{equation}
where each partial derivative \( \frac{\partial C}{\partial a_i} \) is estimated by perturbing the corresponding coefficient \( a_i \) slightly and observing the change in the cost function.

\textbf{Update Rule:} The coefficients are updated using the gradient descent method:
\begin{equation}
\mathbf{a}_{\text{new}} = \mathbf{a}_{\text{old}} - \eta \nabla C(\mathbf{a}_{\text{old}}),
\end{equation}
where $\eta$ is the learning rate, a parameter that controls the step size of the adjustments in the coefficient values.

This iterative process is repeated until a certain number of iterations being reached. The adjustments to the coefficients aim to increasingly align the reconstructed beam's amplitude with that of the target beam, thus optimizing the beam shaping process via maximizing the correlation coefficient. For the reconstructed image in Fig.~\ref{fig:quartic_intensty}(a), $ \eta = 0.05$, 200 iterations have been used leading to a correlation of $\SI{98.5}{\%}$ in the region of interest. Same algorithm has been used to shape the Gaussian into the inverse quartic shape from Fig.~\ref{fig:inverse_quartic}(a), however, here a constant initial phase was enough to yield good results.

Nevertheless, compared to Gaussian beams, Laguerre-Gauss modes exhibit increased sensitivity to phase shift distortion~\cite{boruah2006susceptibility}. In shaping the beam depicted in Fig.~\ref{fig:quartic_intensty}(a), only defocus and higher order spherical aberrations represented by Zernike polynomials have been utilized. However, this algorithm's capabilities extend further; by incorporating also non-radially symmetric Zernike polynomials, it can effectively counteract a broader range of aberrations acquired by the laser beam along its path to the IP~\cite{love1997wave}. This would be the case in experimental setups, where the simulated reconstructed amplitude squared would correspond to an image captured by a CMOS sensor. The choice of the parameters used for the simulation in Fig.~\ref{fig:quartic_intensty} allow for an experimental implementation by using an SLM to display the phase mask calculated in Fig.~\ref{fig:quartic phase shift}(c).

\section{\label{sec:C} Electron probe shape evaluation}

In the electron probe evaluation we include also the energy spread of electrons. The spatial intensity profile of the electron probe having an energy spread $\Delta E$ after passing through an electron lens and propagating to the focal spot can be written as an incoherent sum over electron energies~\cite{haider2000upper}:

\begin{equation}
    \begin{split}
        I(x_2,y_2,E_i) &= \int_{-\infty}^\infty dE_i \, |\psi(x_2,y_2, E_i)|^{2} \frac{1}{\sqrt{2\pi \sigma^2_E}} \times \exp{\left(\frac{-(E_i-E_{0})^2}{2\sigma^2_E}\right)},
    \end{split}
    \label{eqn: I_probe}
\end{equation}
where  $E_0$  is the mean electron energy with $\sigma^2_E = \Delta E^2/(8\ln{2})$ the standard deviation squared of the energy distribution, and $E_i$ the actual electron energy value within the distribution.

 The probe wave function $\psi(x_2,y_2, E_i)$ can be expressed as Fourier transform of the phase shifts distortions acquired in the electron lens plane:

\begin{equation}
    \psi(x_2,y_2, E_i)= FFT(e^{i\chi(\theta,E_i)}),
    \label{eqn: Field_probe 2}
\end{equation}
whereas the aberration function from Eq.~\ref{eq: Aberration function wave} turns into:

\begin{equation}
    \chi(\theta,E_i)= \frac{\pi}{2\lambda_e(E_i)}(C_s\theta^4-2\Delta z'\theta^2),
    \label{eqn: Aberration function}
\end{equation}
with the modified defocus due to the electron energy spread $\Delta z' = \Delta z +C_c\frac{E_i-E_0}{E_0}$, where $C_c$ is the chromatic aberration coefficient.

For weak lenses $f = \SI{8}{mm}$, $C_c \sim f$ ~\cite{reimer2013transmission}, $C_s \sim \SI{80}{mm}$, $\Delta z = 0$, $\theta_{max} =\SI{8.1}{mrad}$, $E_0 = \SI{30}{keV}$ and using field emission gun with $\Delta E = \SI{0.5}{eV}$ we simulate the probe shapes in Fig.\ref{fig:Probe shape}.

\subsubsection*{Misalignment effect of the aberration correction phase plate} 
In this extended model, we also consider the spatial misalignment of the aberration correction phase, implemented as a phase shift displaced in $\nu = 45$ different radially symmetric directions. This modification reflects the possible misalignment of the laser beam with respect to the electron beam and its effect on the probe shape. The misalignment could arise from the vibration of the laser with respect to the electron beam.

The extended wave function from Eq.~\ref{eqn: Field_probe 2}, accounting for phase misalignment, is modified to:

\begin{equation}
    \begin{split}
    \psi_{\text{misaligned}}(x_2, y_2, E_i) &= \sum_{j=1}^{\nu} FFT \biggl[ T_f \cdot \exp \biggl( i \biggl( \chi(\theta, E_i) + \chi_{\text{misaligned}}^{(j)}(\theta, E_i) \biggr) \biggr) \biggr]
    \end{split}
    \label{eqn: Field_probe_misaligned}
\end{equation}

where:

\begin{equation}
    \chi_{\text{misaligned}}^{(j)}(\theta, E_i) = \chi_c(\theta, E_i) \text{ shifted by } [\Delta y_j, \Delta x_j],
\end{equation}
with the correction function $\chi_c(\theta, E_i)$ given by:

\begin{equation}
    \chi_c(\theta, E_i) = -\frac{\pi}{2\lambda_e(E_i)}(C_s\theta^4).
    \label{eqn: Aberration function w/o defocus}
\end{equation}

\begin{figure}[h]
   \centering      
   \includegraphics[width=12.0cm]{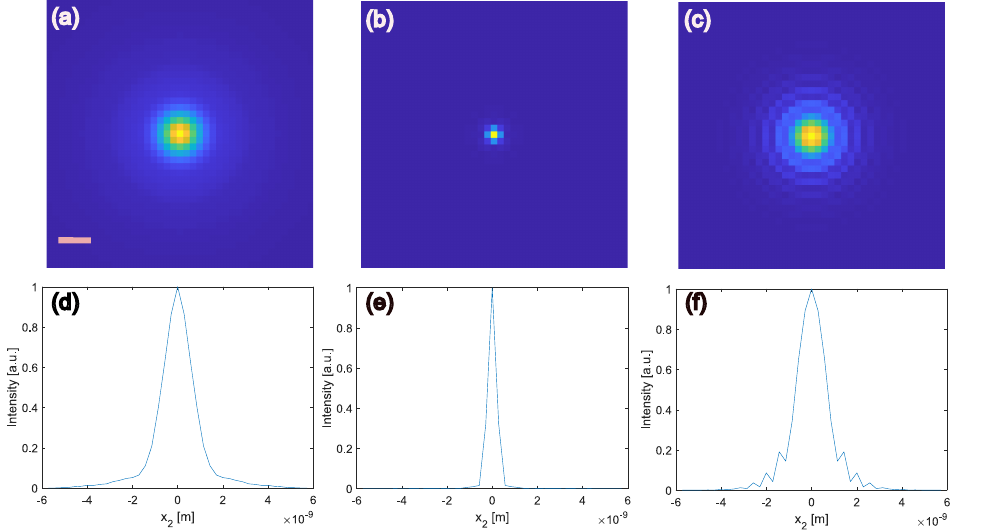}
   \caption{Theoretical calculation of electron probe shapes: Electron focal spot analysis with varying spherical aberration coefficients. Panel (a) shows the electron focal spot (scale bar \SI{1.4}{nm}) with a spherical aberration coefficient $C_s = \SI{80}{mm}$, while panel (b) depicts the electron focal spot without spherical aberration. Panels (d) and (e) illustrate the intensity profiles along the center of the focal spots for panels (a) and (b), respectively. (c) and (f) images are calculated using Eq.~\ref{eqn: Field_probe_misaligned} with displacement distance $d = 1.5\%$ of the correction phase with respect to the diameter of the electron beam in the correction plane.}
   \label{fig:Probe shape}
\end{figure}
The shift components for the $j$-th direction are calculated as:

\begin{equation}
    \begin{split}
    \Delta x_j &= d \cos(\phi_j), \\
    \Delta y_j &= d \sin(\phi_j),
    \end{split}
\end{equation}

where $d$ is the shift displacement distance and $\phi_j = \frac{2\pi j}{\nu}$ is the angle for the $j$-th direction.

The results shown in Fig.~\ref{fig:Probe shape}(c) highlight the importance of the positional stability of the correction phase pattern with respect to the aberrated electron beam.

We conclude that for effective improvement of the probe resolution, the displacement $d$ should be less than $1.5\%$ of $2\cdot\rho_{\text{IP}}$. The degradation in resolution can be attributed to the introduction of additional coma, which is incoherently summed across all $\nu$ radially symmetric directions of the electron beam due to the misalignment of the correction phase.

\begin{backmatter}
\bmsection{Funding}
The authors acknowledge funding from the Czech Science Foundation (project 22-13001K), Charles University (SVV-2023-260720, PRIMUS/19/SCI/05) and the European Union (ERC, eWaveShaper, 101039339). Views and opinions expressed are however those of the author(s) only and do not necessarily reflect those of the European Union or the European Research Council Executive Agency. Neither the European Union nor the granting authority can be held responsible
for them. This work was supported by TERAFIT project No. \text{CZ}.02.01.01/00/22\_008/0004594 funded by OP JAK, call Excellent Research.

\bmsection{Disclosures}
The authors declare no conflicts of interest.

\bmsection{Data availability Statement} Data underlying the results presented in this paper are not publicly available at this time but may be obtained from the authors upon reasonable request.

\end{backmatter}

\bibliography{sample}

\providecommand{\noopsort}[1]{}\providecommand{\singleletter}[1]{#1}
\begin{thebibliography}{10}
\newcommand{\enquote}[1]{``#1''}

\bibitem{zewail2010four}
A.~H. Zewail, \enquote{Four-dimensional electron microscopy,} {\protect\JournalTitle{science}} \textbf{328}, 187--193 (2010).

\bibitem{arbouet2018ultrafast}
A.~Arbouet, G.~M. Caruso, and F.~Houdellier, \enquote{Ultrafast transmission electron microscopy: historical development, instrumentation, and applications,} {\protect\JournalTitle{Advances in imaging and electron physics}} \textbf{207}, 1--72 (2018).

\bibitem{montgomery_leonhardt_roehling_2021}
E.~Montgomery, D.~Leonhardt, and J.~Roehling, \enquote{Ultrafast transmission electron microscopy: Techniques and applications,} {\protect\JournalTitle{Microscopy Today}} \textbf{29}, 46–54 (2021).

\bibitem{morimoto2023attosecond}
Y.~Morimoto, \enquote{Attosecond electron-beam technology: a review of recent progress,} {\protect\JournalTitle{Microscopy}} \textbf{72}, 2--17 (2023).

\bibitem{feist2017ultrafast}
A.~Feist, N.~Bach, N.~R. da~Silva, \emph{et~al.}, \enquote{Ultrafast transmission electron microscopy using a laser-driven field emitter: Femtosecond resolution with a high coherence electron beam,} {\protect\JournalTitle{Ultramicroscopy}} \textbf{176}, 63--73 (2017).

\bibitem{houdellier2018development}
F.~Houdellier, G.~M. Caruso, S.~Weber, \emph{et~al.}, \enquote{Development of a high brightness ultrafast transmission electron microscope based on a laser-driven cold field emission source,} {\protect\JournalTitle{Ultramicroscopy}} \textbf{186}, 128--138 (2018).

\bibitem{zhu2020development}
C.~Zhu, D.~Zheng, H.~Wang, \emph{et~al.}, \enquote{Development of analytical ultrafast transmission electron microscopy based on laser-driven schottky field emission,} {\protect\JournalTitle{Ultramicroscopy}} \textbf{209}, 112887 (2020).

\bibitem{haider1998spherical}
M.~Haider, H.~Rose, S.~Uhlemann, \emph{et~al.}, \enquote{A spherical-aberration-corrected 200 kv transmission electron microscope,} {\protect\JournalTitle{Ultramicroscopy}} \textbf{75}, 53--60 (1998).

\bibitem{muller2006advancing}
H.~M{\"u}ller, S.~Uhlemann, P.~Hartel, and M.~Haider, \enquote{Advancing the hexapole cs-corrector for the scanning transmission electron microscope,} {\protect\JournalTitle{Microscopy and Microanalysis}} \textbf{12}, 442--455 (2006).

\bibitem{rose2009historical}
H.~H. Rose, \enquote{Historical aspects of aberration correction,} {\protect\JournalTitle{Journal of electron microscopy}} \textbf{58}, 77--85 (2009).

\bibitem{krivanek1999towards}
O.~Krivanek, N.~Dellby, and A.~Lupini, \enquote{Towards sub-{\aa} electron beams,} {\protect\JournalTitle{Ultramicroscopy}} \textbf{78}, 1--11 (1999).

\bibitem{shiloh2018spherical}
R.~Shiloh, R.~Remez, P.-H. Lu, \emph{et~al.}, \enquote{Spherical aberration correction in a scanning transmission electron microscope using a sculpted thin film,} {\protect\JournalTitle{Ultramicroscopy}} \textbf{189}, 46--53 (2018).

\bibitem{roitman2021shaping}
D.~Roitman, R.~Shiloh, P.-H. Lu, \emph{et~al.}, \enquote{Shaping of electron beams using sculpted thin films,} {\protect\JournalTitle{ACS photonics}} \textbf{8}, 3394--3405 (2021).

\bibitem{maurer2011spatial}
C.~Maurer, A.~Jesacher, S.~Bernet, and M.~Ritsch-Marte, \enquote{What spatial light modulators can do for optical microscopy,} {\protect\JournalTitle{Laser \& Photonics Reviews}} \textbf{5}, 81--101 (2011).

\bibitem{Rubinsztein_Dunlop_2016}
H.~Rubinsztein-Dunlop, A.~Forbes, M.~V. Berry, \emph{et~al.}, \enquote{Roadmap on structured light,} {\protect\JournalTitle{Journal of Optics}} \textbf{19}, 013001 (2016).

\bibitem{hampson2021adaptive}
K.~M. Hampson, R.~Turcotte, D.~T. Miller, \emph{et~al.}, \enquote{Adaptive optics for high-resolution imaging,} {\protect\JournalTitle{Nature Reviews Methods Primers}} \textbf{1}, 1--26 (2021).

\bibitem{VERBEECK201858}
J.~Verbeeck, A.~Béché, K.~Müller-Caspary, \emph{et~al.}, \enquote{Demonstration of a 2×2 programmable phase plate for electrons,} {\protect\JournalTitle{Ultramicroscopy}} \textbf{190}, 58--65 (2018).

\bibitem{ibanez2022}
F.~V. Ibáñez, A.~Béché, and J.~Verbeeck, \enquote{Can a programmable phase plate serve as an aberration corrector in the transmission electron microscope (tem)?}  (2022).

\bibitem{ribet2023design}
S.~M. Ribet, S.~E. Zeltmann, K.~C. Bustillo, \emph{et~al.}, \enquote{Design of electrostatic aberration correctors for scanning transmission electron microscopy,} {\protect\JournalTitle{Microscopy and Microanalysis}} \textbf{29}, 1950--1960 (2023).

\bibitem{yu2023quantum}
C.-P. Yu, F.~Vega~Iba{\~n}ez, A.~B{\'e}ch{\'e}, and J.~Verbeeck, \enquote{Quantum wavefront shaping with a 48-element programmable phase plate for electrons,} {\protect\JournalTitle{SciPost Physics}} \textbf{15}, 223 (2023).

\bibitem{grillo2014}
V.~Grillo, E.~Karimi, G.~C. Gazzadi, \emph{et~al.}, \enquote{Generation of nondiffracting electron bessel beams,} {\protect\JournalTitle{Phys. Rev. X}} \textbf{4}, 011013 (2014).

\bibitem{barwick2009photon}
B.~Barwick, D.~J. Flannigan, and A.~H. Zewail, \enquote{Photon-induced near-field electron microscopy,} {\protect\JournalTitle{Nature}} \textbf{462}, 902--906 (2009).

\bibitem{vanacore2018attosecond}
G.~M. Vanacore, I.~Madan, G.~Berruto, \emph{et~al.}, \enquote{Attosecond coherent control of free-electron wave functions using semi-infinite light fields,} {\protect\JournalTitle{Nature communications}} \textbf{9}, 2694 (2018).

\bibitem{vanacore2019ultrafast}
G.~M. Vanacore, G.~Berruto, I.~Madan, \emph{et~al.}, \enquote{Ultrafast generation and control of an electron vortex beam via chiral plasmonic near fields,} {\protect\JournalTitle{Nature materials}} \textbf{18}, 573--579 (2019).

\bibitem{konevcna2020electron}
A.~Kone{\v{c}}n{\'a} and F.~J.~G. de~Abajo, \enquote{Electron beam aberration correction using optical near fields,} {\protect\JournalTitle{Physical Review Letters}} \textbf{125}, 030801 (2020).

\bibitem{ben2021shaping}
A.~Ben~Hayun, O.~Reinhardt, J.~Nemirovsky, \emph{et~al.}, \enquote{Shaping quantum photonic states using free electrons,} {\protect\JournalTitle{Science Advances}} \textbf{7}, eabe4270 (2021).

\bibitem{henke2021integrated}
J.-W. Henke, A.~S. Raja, A.~Feist, \emph{et~al.}, \enquote{Integrated photonics enables continuous-beam electron phase modulation,} {\protect\JournalTitle{Nature}} \textbf{600}, 653--658 (2021).

\bibitem{dahan2021imprinting}
R.~Dahan, A.~Gorlach, U.~Haeusler, \emph{et~al.}, \enquote{Imprinting the quantum statistics of photons on free electrons,} {\protect\JournalTitle{Science}} \textbf{373}, eabj7128 (2021).

\bibitem{feist2022cavity}
A.~Feist, G.~Huang, G.~Arend, \emph{et~al.}, \enquote{Cavity-mediated electron-photon pairs,} {\protect\JournalTitle{Science}} \textbf{377}, 777--780 (2022).

\bibitem{madan2022ultrafast}
I.~Madan, V.~Leccese, A.~Mazur, \emph{et~al.}, \enquote{Ultrafast transverse modulation of free electrons by interaction with shaped optical fields,} {\protect\JournalTitle{ACS photonics}} \textbf{9}, 3215--3224 (2022).

\bibitem{tsesses2023tunable}
S.~Tsesses, R.~Dahan, K.~Wang, \emph{et~al.}, \enquote{Tunable photon-induced spatial modulation of free electrons,} {\protect\JournalTitle{Nature Materials}} \textbf{22}, 345--352 (2023).

\bibitem{garcia2023spatiotemporal}
F.~J. Garc{\'\i}a~de Abajo and C.~Ropers, \enquote{Spatiotemporal electron beam focusing through parallel interactions with shaped optical fields,} {\protect\JournalTitle{Physical Review Letters}} \textbf{130}, 246901 (2023).

\bibitem{Bucksbaum1988}
P.~H. Bucksbaum, D.~W. Schumacher, and M.~Bashkansky, \enquote{{High-intensity kapitza-dirac effect},} {\protect\JournalTitle{Physical Review Letters}} \textbf{61}, 1182--1185 (1988).

\bibitem{freimund2001observation}
D.~L. Freimund, K.~Aflatooni, and H.~Batelaan, \enquote{Observation of the kapitza--dirac effect,} {\protect\JournalTitle{Nature}} \textbf{413}, 142--143 (2001).

\bibitem{dwyer2006femtosecond}
J.~R. Dwyer, C.~T. Hebeisen, R.~Ernstorfer, \emph{et~al.}, \enquote{Femtosecond electron diffraction:‘making the molecular movie’,} {\protect\JournalTitle{Philosophical Transactions of the Royal Society A: Mathematical, Physical and Engineering Sciences}} \textbf{364}, 741--778 (2006).

\bibitem{Kozak2018}
M.~Koz{\'{a}}k, N.~Sch{\"{o}}nenberger, and P.~Hommelhoff, \enquote{{Ponderomotive Generation and Detection of Attosecond Free-Electron Pulse Trains},} {\protect\JournalTitle{Physical Review Letters}} \textbf{120}, 103203 (2018).

\bibitem{schwartz2019laser}
O.~Schwartz, J.~J. Axelrod, S.~L. Campbell, \emph{et~al.}, \enquote{Laser phase plate for transmission electron microscopy,} {\protect\JournalTitle{Nature methods}} \textbf{16}, 1016--1020 (2019).

\bibitem{axelrod2020observation}
J.~J. Axelrod, S.~L. Campbell, O.~Schwartz, \emph{et~al.}, \enquote{Observation of the relativistic reversal of the ponderomotive potential,} {\protect\JournalTitle{Physical review letters}} \textbf{124}, 174801 (2020).

\bibitem{chirita2022transverse}
M.~C. Chirita~Mihaila, P.~Weber, M.~Schneller, \emph{et~al.}, \enquote{Transverse electron-beam shaping with light,} {\protect\JournalTitle{Physical Review X}} \textbf{12}, 031043 (2022).

\bibitem{tsarev2023nonlinear}
M.~Tsarev, J.~W. Thurner, and P.~Baum, \enquote{Nonlinear-optical quantum control of free-electron matter waves,} {\protect\JournalTitle{Nature Physics}} \textbf{19}, 1350--1354 (2023).

\bibitem{ebel2023inelastic}
S.~Ebel and N.~Talebi, \enquote{Inelastic electron scattering at a single-beam structured light wave,} {\protect\JournalTitle{Communications Physics}} \textbf{6}, 179 (2023).

\bibitem{streshkova2024electron}
N.~L. Streshkova, P.~Koutensk{\`y}, and M.~Koz{\'a}k, \enquote{Electron vortex beams for chirality probing at the nanoscale,} {\protect\JournalTitle{Physical Review Applied}} \textbf{22}, 054017 (2024).

\bibitem{streshkova2024monochromatization}
N.~L. Streshkova, P.~Koutensk{\`y}, T.~Novotn{\`y}, and M.~Koz{\'a}k, \enquote{Monochromatization of electron beams with spatially and temporally modulated optical fields,} {\protect\JournalTitle{Physical Review Letters}} \textbf{133}, 213801 (2024).

\bibitem{velasco2024free}
C.~I. Velasco and F.~de~Abajo, \enquote{Free-space optical modulation of free electrons in the continuous-wave regime,} {\protect\JournalTitle{arXiv preprint arXiv:2412.03410}}  (2024).

\bibitem{de2021optical}
F.~J.~G. de~Abajo and A.~Kone{\v{c}}n{\'a}, \enquote{Optical modulation of electron beams in free space,} {\protect\JournalTitle{Physical Review Letters}} \textbf{126}, 123901 (2021).

\bibitem{uesugi2021electron}
Y.~Uesugi, Y.~Kozawa, and S.~Sato, \enquote{Electron round lenses with negative spherical aberration by a tightly focused cylindrically polarized light beam,} {\protect\JournalTitle{Physical Review Applied}} \textbf{16}, L011002 (2021).

\bibitem{uesugi2022properties}
Y.~Uesugi, Y.~Kozawa, and S.~Sato, \enquote{Properties of electron lenses produced by ponderomotive potential with bessel and laguerre--gaussian beams,} {\protect\JournalTitle{Journal of Optics}} \textbf{24}, 054013 (2022).

\bibitem{saxton2000new}
W.~Saxton, \enquote{A new way of measuring microscope aberrations,} {\protect\JournalTitle{Ultramicroscopy}} \textbf{81}, 41--45 (2000).

\bibitem{cowley1979adjustment}
J.~Cowley, \enquote{Adjustment of a stem instrument by use of shadow images,} {\protect\JournalTitle{Ultramicroscopy}} \textbf{4}, 413--418 (1979).

\bibitem{lin1986calibration}
J.~Lin and J.~Cowley, \enquote{Calibration of the operating parameters for an hb5 stem instrument,} {\protect\JournalTitle{Ultramicroscopy}} \textbf{19}, 31--42 (1986).

\bibitem{james1999practical}
E.~James and N.~Browning, \enquote{Practical aspects of atomic resolution imaging and analysis in stem,} {\protect\JournalTitle{Ultramicroscopy}} \textbf{78}, 125--139 (1999).

\bibitem{liu2005scanning}
J.~Liu, \enquote{Scanning transmission electron microscopy and its application to the study of nanoparticles and nanoparticle systems,} {\protect\JournalTitle{Journal of electron microscopy}} \textbf{54}, 251--278 (2005).

\bibitem{sawada2008measurement}
H.~Sawada, T.~Sannomiya, F.~Hosokawa, \emph{et~al.}, \enquote{Measurement method of aberration from ronchigram by autocorrelation function,} {\protect\JournalTitle{Ultramicroscopy}} \textbf{108}, 1467--1475 (2008).

\bibitem{spangenberg1942some}
K.~Spangenberg and L.~M. Field, \enquote{Some simplified methods of determining the optical characteristics of electron lenses,} {\protect\JournalTitle{Proceedings of the IRE}} \textbf{30}, 138--144 (1942).

\bibitem{rempfer1985unipotential}
G.~F. Rempfer, \enquote{Unipotential electrostatic lenses: Paraxial properties and aberrations of focal length and focal point,} {\protect\JournalTitle{Journal of applied physics}} \textbf{57}, 2385--2401 (1985).

\bibitem{rempfer1997simultaneous}
G.~F. Rempfer, D.~M. Desloge, W.~P. Skoczylas, and O.~H. Griffith, \enquote{Simultaneous correction of spherical and chromatic aberrations with an electron mirror: an electron optical achromat,} {\protect\JournalTitle{Microscopy and Microanalysis}} \textbf{3}, 14--27 (1997).

\bibitem{scherzer1949theoretical}
O.~Scherzer, \enquote{The theoretical resolution limit of the electron microscope,} {\protect\JournalTitle{Journal of Applied Physics}} \textbf{20}, 20--29 (1949).

\bibitem{PhysRevLett.126.123901}
F.~J. Garc\'{\i}a~de Abajo and A.~Kone\ifmmode~\check{c}\else \v{c}\fi{}n\'a, \enquote{Optical modulation of electron beams in free space,} {\protect\JournalTitle{Phys. Rev. Lett.}} \textbf{126}, 123901 (2021).

\bibitem{ChiritaMihaila2022}
M.~C. Chirita~Mihaila, \enquote{Electron beam shaping with light,} Phd diss., University of Vienna (2022). See pp. 39.

\bibitem{uhlemann2013thermal}
S.~Uhlemann, H.~M{\"u}ller, P.~Hartel, \emph{et~al.}, \enquote{Thermal magnetic field noise limits resolution in transmission electron microscopy,} {\protect\JournalTitle{Physical review letters}} \textbf{111}, 046101 (2013).

\bibitem{xie2015phase}
G.~Xie, Y.~Ren, H.~Huang, \emph{et~al.}, \enquote{Phase correction for a distorted orbital angular momentum beam using a zernike polynomials-based stochastic-parallel-gradient-descent algorithm,} {\protect\JournalTitle{Optics letters}} \textbf{40}, 1197--1200 (2015).

\bibitem{gerchberg1972practical}
R.~W. Gerchberg, \enquote{A practical algorithm for the determination of phase from image and diffraction plane pictures,} {\protect\JournalTitle{Optik}} \textbf{35}, 237--246 (1972).

\bibitem{nocedal1999numerical}
J.~Nocedal and S.~J. Wright, \emph{Numerical optimization} (Springer, 1999).

\bibitem{tan2019silicon}
S.~Tan, Z.~Zhao, K.~Urbanek, \emph{et~al.}, \enquote{Silicon nitride waveguide as a power delivery component for on-chip dielectric laser accelerators,} {\protect\JournalTitle{Optics letters}} \textbf{44}, 335--338 (2019).

\bibitem{van2010compression}
T.~Van~Oudheusden, P.~Pasmans, S.~Van Der~Geer, \emph{et~al.}, \enquote{Compression of subrelativistic space-charge-dominated electron bunches for single-shot femtosecond electron diffraction,} {\protect\JournalTitle{Physical review letters}} \textbf{105}, 264801 (2010).

\bibitem{zhou2017femtosecond}
F.~Zhou, J.~Williams, and C.-Y. Ruan, \enquote{Femtosecond electron spectroscopy in an electron microscope with high brightness beams,} {\protect\JournalTitle{Chemical Physics Letters}} \textbf{683}, 488--494 (2017).

\bibitem{williams2017active}
J.~Williams, F.~Zhou, T.~Sun, \emph{et~al.}, \enquote{Active control of bright electron beams with rf optics for femtosecond microscopy,} {\protect\JournalTitle{Structural Dynamics}} \textbf{4} (2017).

\bibitem{berruto2018laser}
G.~Berruto, I.~Madan, Y.~Murooka, \emph{et~al.}, \enquote{Laser-induced skyrmion writing and erasing in an ultrafast cryo-lorentz transmission electron microscope,} {\protect\JournalTitle{Physical review letters}} \textbf{120}, 117201 (2018).

\bibitem{rubiano2018nanoscale}
N.~Rubiano~da Silva, M.~M{\"o}ller, A.~Feist, \emph{et~al.}, \enquote{Nanoscale mapping of ultrafast magnetization dynamics with femtosecond lorentz microscopy,} {\protect\JournalTitle{Physical Review X}} \textbf{8}, 031052 (2018).

\bibitem{kim2008imaging}
J.~S. Kim, T.~LaGrange, B.~W. Reed, \emph{et~al.}, \enquote{Imaging of transient structures using nanosecond in situ tem,} {\protect\JournalTitle{Science}} \textbf{321}, 1472--1475 (2008).

\bibitem{du2024development}
D.~X. Du, A.~C. Bartnik, C.~J. Duncan, \emph{et~al.}, \enquote{Development of an ultrafast pulsed ponderomotive phase plate for cryo-electron tomography,} {\protect\JournalTitle{bioRxiv}} pp. 2024--03 (2024).

\bibitem{MATLAB}
T.~M. Inc., \enquote{Matlab version: 9.13.0 (r2022b),}  (2022).

\bibitem{voelz2011computational}
D.~G. Voelz, \emph{Computational fourier optics: a MATLAB tutorial}, vol. 534 (SPIE press Bellingham, Washington, 2011).

\bibitem{lakshminarayanan2011zernike}
V.~Lakshminarayanan and A.~Fleck, \enquote{Zernike polynomials: a guide,} {\protect\JournalTitle{Journal of Modern Optics}} \textbf{58}, 545--561 (2011).

\bibitem{monmayrant2004new}
A.~Monmayrant and B.~Chatel, \enquote{New phase and amplitude high resolution pulse shaper,} {\protect\JournalTitle{Review of Scientific Instruments}} \textbf{75}, 2668--2671 (2004).

\bibitem{boruah2006susceptibility}
B.~R. Boruah and M.~Neil, \enquote{Susceptibility to and correction of azimuthal aberrations in singular light beams,} {\protect\JournalTitle{Optics express}} \textbf{14}, 10377--10385 (2006).

\bibitem{love1997wave}
G.~D. Love, \enquote{Wave-front correction and production of zernike modes with a liquid-crystal spatial light modulator,} {\protect\JournalTitle{Applied optics}} \textbf{36}, 1517--1524 (1997).

\bibitem{haider2000upper}
M.~Haider, S.~Uhlemann, and J.~Zach, \enquote{Upper limits for the residual aberrations of a high-resolution aberration-corrected stem,} {\protect\JournalTitle{Ultramicroscopy}} \textbf{81}, 163--175 (2000).

\bibitem{reimer2013transmission}
L.~Reimer, \emph{Transmission electron microscopy: physics of image formation and microanalysis}, vol.~36 (Springer, 2013). See pp. 39.

\end{thebibliography}

\end{document}